\documentclass[10pt,letterpaper]{article}
\usepackage[top=0.85in,left=2.75in,footskip=0.75in]{geometry}

\usepackage{amsmath,amssymb}
\usepackage{ amsfonts, amscd} 
\usepackage{amsthm} 
\usepackage[mathscr]{eucal}
\usepackage{changepage} 

\usepackage[utf8x]{inputenc}

\usepackage{textcomp,marvosym}

\usepackage{cite}

\usepackage{nameref,hyperref}

\usepackage[right]{lineno}

\usepackage{microtype}
\DisableLigatures[f]{encoding = *, family = * }

\usepackage[table]{xcolor}

\usepackage{array}

\newcolumntype{+}{!{\vrule width 2pt}}

\newlength\savedwidth

\newcommand\thickhline{\noalign{\global\savedwidth\arrayrulewidth\global\arrayrulewidth 2pt}%
\hline
\noalign{\global\arrayrulewidth\savedwidth}}

\raggedright
\usepackage[aboveskip=1pt,labelfont=bf,labelsep=period,justification=raggedright,singlelinecheck=off]{caption}

\makeatletter
\renewcommand{\@biblabel}[1]{\quad#1.}
\makeatother

\usepackage{fancyhdr,graphicx, lastpage}

\usepackage{epstopdf}
\fancyheadoffset[L]{2.25in}
\fancyfootoffset[L]{2.25in}
\begin{document}
\vspace*{0.2in}

\begin{flushleft}
{\Large
\textbf\newline{Long Title: When does humoral memory enhance infection?} 
}
\newline
Short Title: When does humoral memory enhance infection?
\newline
\\
Ariel Nikas\textsuperscript{1\P},
Hasan Ahmed\textsuperscript{2\P},
Mia R. Moore\textsuperscript{3},
Veronika I. Zarnitsyna\textsuperscript{1},
Rustom Antia\textsuperscript{2*}
\\
\bigskip
\textbf{1} Department of Microbiology and Immunology, Emory University School of Medicine, Atlanta, Georgia, USA
\\
\textbf{2} Department of Biology, Emory University, Atlanta, Georgia, USA
\\
\textbf{3} Fred Hutchinson Cancer Research Center, Seattle, Washington, USA
\\

\bigskip

%

* Corresponding author email: rantia@emory.edu
\bigskip

\P These authors contributed equally to this work
\end{flushleft}
\newpage
\section*{Abstract}
Antibodies and humoral memory are key components of the adaptive immune system. We consider and computationally model mechanisms by which humoral memory present at baseline might instead increase infection load; we refer to this effect as EI-HM (enhancement of infection by humoral memory). We first consider antibody dependent enhancement (ADE) in which antibody enhances the growth of the pathogen, typically a virus, and typically at intermediate `Goldilocks' levels of antibody. Our ADE model reproduces ADE  {\it in vitro} and enhancement of infection {\it in vivo} from passive antibody transfer. But notably the simplest implementation of our ADE model never results in EI-HM. Adding complexity, by making the cross-reactive antibody much less neutralizing than the  {\it de novo} generated antibody or by including a sufficiently strong non-antibody immune response, allows for ADE-mediated EI-HM. We next consider the possibility that cross-reactive memory causes EI-HM by crowding out a possibly superior  {\it de novo} immune response. We show that, even without ADE,  EI-HM can occur when the cross-reactive response is both less potent and `directly' (i.e. independently of infection load) suppressive with regard to the  {\it de novo} response. In this case adding a non-antibody immune response to our computational model greatly reduces or completely eliminates EI-HM, which suggests that `crowding out' is unlikely to cause substantial EI-HM. Hence, our results provide examples in which simple models give qualitatively opposite results compared to models with plausible complexity. Our results may be helpful in interpreting and reconciling disparate experimental findings, especially from dengue, and for vaccination.

\section*{Author summary}
Humoral memory, generated by infection with a pathogen, can protect against subsequent infection by the same and related pathogens. We consider situations in which humoral memory is counterproductive and instead enhances infection by related pathogens. Numerous experiments have shown that the addition of binding antibody makes certain viruses more, rather than less, infectious; typically high levels of antibody are still protective, meaning that infectivity is maximized at intermediate `Goldilocks' levels of antibody. Additionally, we consider the situation in which cross-reactive humoral memory dominates relative to the {\it de novo} response. Memory dominance has been documented for influenza infections, but whether it is harmful or not is unclear. We show computationally that both ADE and a certain type of competition between the cross-reactive and {\it de novo} responses translate into enhancement of infection in certain circumstances but not in others. We discuss the implications of our research for dengue infection, a common mosquito-transmitted viral infection, and vaccination.

\section*{Introduction}
Antibody is a key component of the adaptive immune system. Antibodies bind to viruses, bacteria, and other microbes and thereby inhibit microbe function and aid in microbe clearance. For certain infections, antibody levels are considered the key correlate of protection ~\cite{plotkin2010correlates}, meaning that increasing antibody levels are associated with protection from infection or disease.

Paradoxically for certain viral infections, antibody can instead enhance infection - an effect known as antibody dependent enhancement (ADE). ADE has now been reported for many viruses including dengue, West Nile, HIV, influenza A, Ebola, rabies, polio, and HSV1 ~\cite{suhrbier2003suppression}. It is generally believed that ADE results when antibody bound to virus increases the ability of that virus to infect certain cells, like macrophages, that possess Fc receptors ~\cite{suhrbier2003suppression}.

In the case of dengue, it has also been reported that maternal antibodies can enhance disease severity in infants ~\cite{kliks1988evidence} ~\cite{chau2008dengue}. Consistent with this observation, transfer of dengue antibodies to dengue-infected monkeys can greatly increase viral load ~\cite{halstead1979vivo} ~\cite{goncalvez2007monoclonal}.

A related concern to ADE is that cross-reactive humoral memory from previous infections or vaccinations may enhance the subsequent infection ~\cite{guzman2016dengue} ~\cite{ferguson2016benefits}. This is particularly a concern for dengue infection which is caused by four serotypes of virus – DENV1, DENV2, DENV3, and DENV4 – that are only ~65\% similar at the amino acid level ~\cite{hahn1988nucleotide}. Certain studies have reported that infection with one serotype of dengue, for example DENV1, may make infection with a second serotype, for example DENV2, more severe probably as a consequence of ADE ~\cite{guzman2000epidemiologic} ~\cite{vaughn2000dengue} ~\cite{katzelnick2017antibody}.

In contrast, the discussion surrounding influenza has focused more on the issue of original antigenic sin ~\cite{henry2017original}. According to the original antigenic sin model, the immune response remains focused on the first influenza strain encountered even after multiple subsequent influenza infections. Original antigenic sin is not necessarily bad, and certain authors have argued that antigenic seniority is a more apt term ~\cite{henry2017original}.


\subsection*{Definitions}
In this paper we use the following definitions:

{\em Antibody dependent enhancement} (ADE) means that viral growth is enhanced at some level of antibody compared to no antibody. Mechanistically, we model this via higher infectivity of virions with some bound antibody.

{\em Enhancement of infection from humoral memory} (EI-HM) means that infection load is greater in the presence of cross-reactive humoral memory than in its absence. While there are many ways to measure infection load, in this paper we consistently use the peak number of infected cells.

{\em Enhancement of infection from passive antibody} (EI-PA) means that infection load is greater when passive antibody – for example maternal antibody – is supplied to a host than in its absence. (We assume that this host has an adaptive immune system and will generate a humoral response of its own.)

{\em Memory dominance} means that the cross-reactive memory response dominates the {\it de novo} immune response.  Memory dominance may be viewed as a proxy for original antigenic sin (see Supplement for further explanation). 

{\em Suppressive memory} means that cross-reactive memory “directly” suppresses the {\it de novo} immune response ignoring any “indirect” effects mediated via infection load. 

\begin{figure}[!h]
\centering
  \includegraphics[scale=0.35]{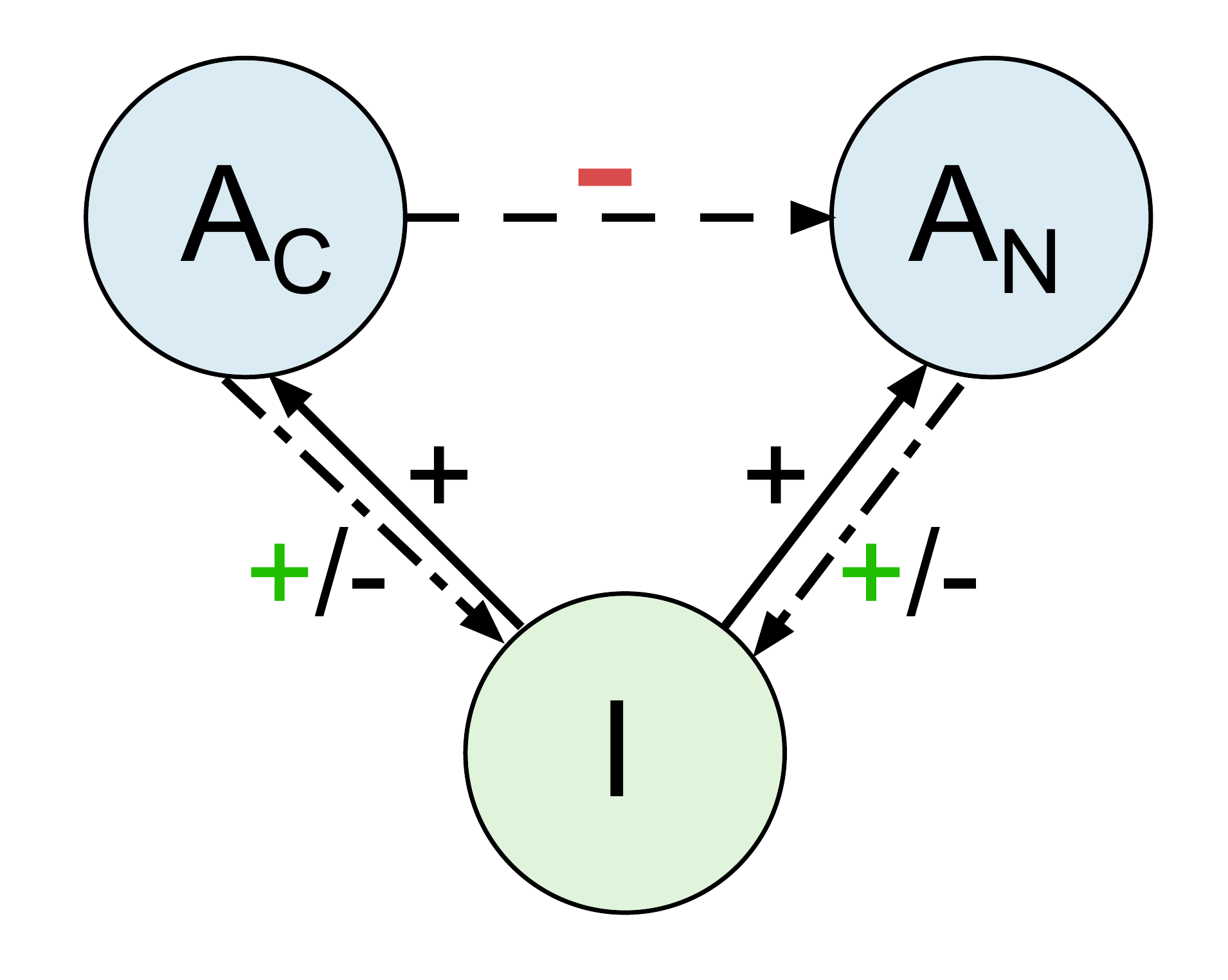}
\caption{{\bf  Conceptual diagram.}
Classically, infected cells ($I$) trigger the growth of antibodies ($A_c$ and $A_N$), and antibodies inhibit infection. In the case of ADE, antibodies can instead enhance infection (shown by green plus signs). In the case of suppressive memory (effect shown by the red minus sign), the cross-reactive response ($A_c$) directly inhibits the response of {\it de novo} antibodies ($A_N$); this is in addition to any indirect effect mediated by infection load.}
\label{fig1A}
\end{figure}

See Fig~\ref{fig1A} for a conceptual diagram of these differences.
\section*{Model and Results}
\subsection*{ADE Model}
To investigate the mechanisms depicted in Fig~\ref{fig1A}, we create a mathematical model of the essential infection components. In this model, infected cells ($I$) produce virus. $V_1$ is virus with little or no bound antibody. $V_2$ is virus with intermediate amounts of bound antibody. Virus with the most antibody bound, $V_3$, is neutralized. We assume that $V_2$ virus is more infectious than $V_1$ virus (i.e. $\beta_2 > \beta_1$). $A_c$ is cross-reactive memory antibody. $A_n$ is the {\it de novo} antibody response. $A_p$ is passive antibody (e.g. maternal antibody). $R$ is the non-antibody (e.g. innate or CD8 T cell) immune response. We assume that only a small fraction of target cells are depleted during the course of infection; hence, we do not explicitly model loss of susceptible cells. We show this model set up via a schematic in ~\nameref{S1_Fig}.


 The mechanisms we consider are described by the following set of ordinary differential equations.

\begin{align}
\dot{I} &= \underbrace{-bI}_{\text{cell death}}+\underbrace{\beta_1V_1+\beta_2V_2}_{\text{infection}}-\underbrace{IR}_{\text{non-Ab clearance}}\\
\dot{V_1} &= \underbrace{I}_{\text{production}} - V_1(\underbrace{a}_{\text{clearance}}+\underbrace{k_{c1}A_c+k_{n1}A_n+k_{p1}A_p}_{\text{binding by antibodies}})\\
\dot{V_2} &= \underbrace{V_1(k_{c1}A_c+k_{n1}A_n+k_{p1}A_p)}_{\text{newly partially bound}} - V_2(\underbrace{a}_{\text{clearance}}+\underbrace{k_{c2}A_c+k_{n2}A_n+k_{p2}A_p)}_{\text{further binding}}\\
\dot{V_3} &= \underbrace{V_2(k_{c2}A_c+k_{n2}A_n+k_{p2}A_p)}_{\text{newly  completely bound}}-\underbrace{aV_3}_{\text{clearance}}\\
\dot{A_c} &= \underbrace{s_c}_{\text{growth}}\underbrace{\chi(I>1)}_{\text{indicator}}A_c\\
\dot{A_n} &= \underbrace{s_n}_{\text{growth}}\underbrace{\chi(I>1)}_{\text{indicator}}A_n\\
\dot{A_p} &= \underbrace{0}_{\text{no growth}} \\
\dot{R} &= \underbrace{s_R}_{\text{growth}}\underbrace{\chi(I>1)}_{\text{indicator}}
\end{align}
Here, $\chi$ is the indicator function. Table~\ref{table1} describes the model parameters. Unless otherwise stated we use the default parameter values shown in the table. Model parameters were chosen to give an asymptotic growth of 1.5 natural logs (ln) per day in the absence of antibody and 2.5 ln per day at maximum ADE and peak infection load at 10 days. For comparison the growth of the yellow fever YFV-17D virus was estimated at 1.6 ln per day \cite{Moore18}. These dynamics are shown in Fig S2-7 for each of the models.
\begin{table}[!ht]
\centering
\caption{
{\bf Parameters in ADE model.}}
\begin{tabular}{| l | l | l |}
\hline
Parameter & Interpretation & Default Value*\\ \thickhline
b & Death rate of infected cells & 1/day \\ \hline
$\beta_1$ & Infectivity of $V_1$ virus & 29 CU/VU/day \\ \hline
$\beta_2 $& Infectivity of $V_2$ virus & 74 CU/VU/day \\ \hline
$a$ & Death rate of virions & 10/day \\ \hline
$k_{c1}, k_{n1}, k_{p1} $& Efficiency of conversion of $V_1$  & 1/AU/day\\ 
& virus to $V_2$ virus by $A_c$, $A_n$, and $A_p$& \\
  & antibody respectively &  \\ \hline
$k_{c2}, k_{n2}, k_{p2} $& Efficiency of neutralization of   & 0.25/AU/day \\ 
& $V_2$ virus by $A_c$, $A_n$, and $A_p$& \\
&   antibody respectively & \\ \hline
$s_{c}, s_{n} $& Growth rate of $A_c$ and $A_n$  & 1/day \\
& antibody respectively& \\  \hline
$s_{R} $& Growth rate of non-antibody  & 0 RU/day\\ 
& immune response& \\ \hline
$I(0) $& Infected cells at time zero & 0 CU \\ \hline
$V_1(0), V_2(0), V_3(0) $& Initial values for virus & 0.1, 0, and 0 VU \\ \hline
$A_c(0), A_n(0), A_p(0) $& Initial values for antibody & 0, 0.036, and 0 AU \\ \hline
$R(0) $& Initial value for non-antibody  & 0 RU \\
& immune response& \\ \hline
\end{tabular}
\begin{flushleft} *CU, VU, AU and RU are units for cells, virus, antibody and non-antibody immunity respectively.
\end{flushleft}
\label{table1}
\end{table}



\subsection*{The model recapitulates{ \em in vitro} ADE}
\begin{figure}[!h]
\centering
  \includegraphics[scale=0.75]{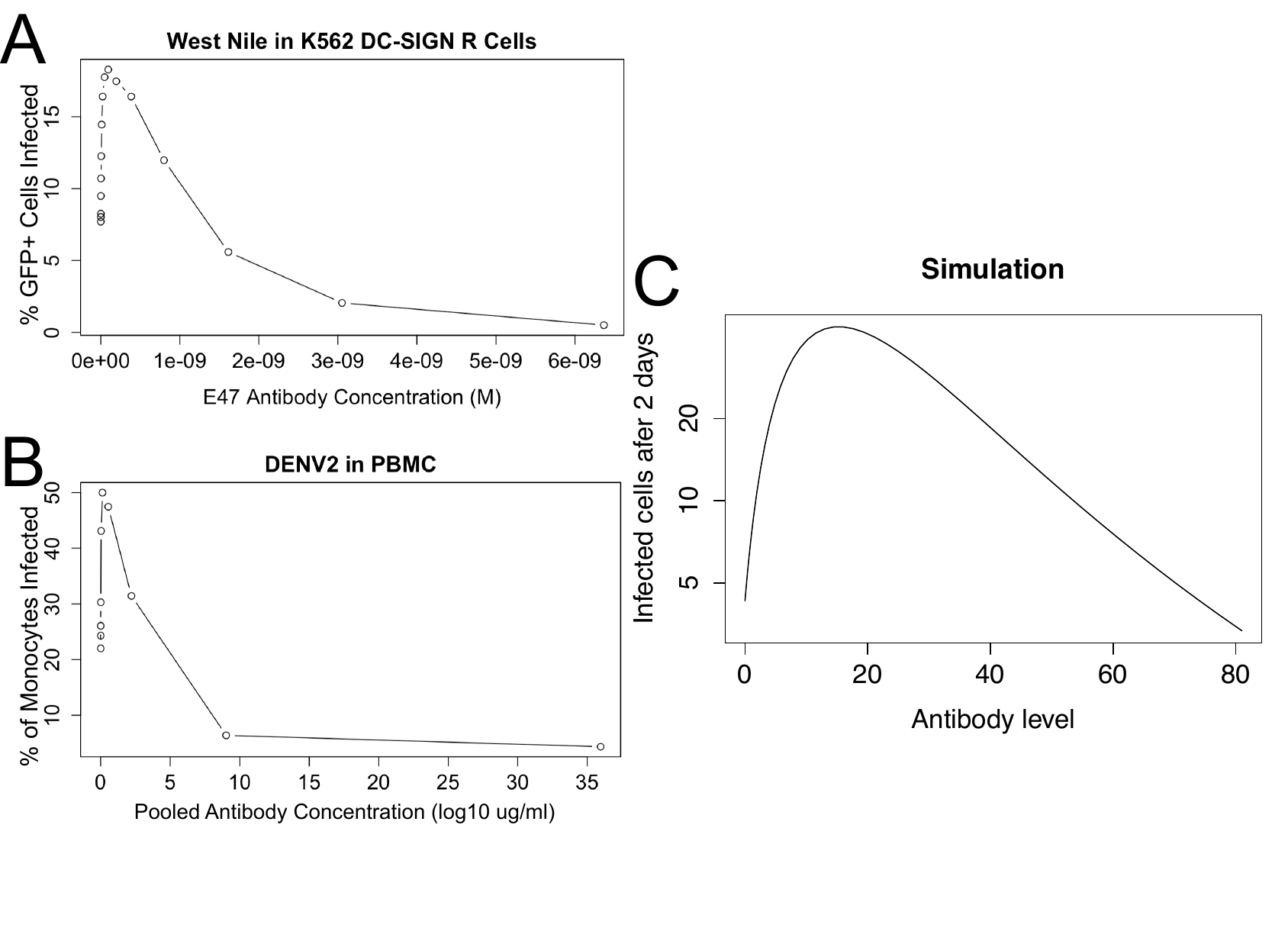}
\caption{{\bf  Experimental and simulated \textbf{\textit{in vitro}} ADE.}
 In Panel A, a single cycle West Nile virus, a flavivirus related to dengue, was grown in K562 DC-SIGN R cells with different molar concentrations of antibody which yielded peak infection load at intermediate concentrations of antibody. Dengue, grown for 24 hours in PBMC with different concentrations of dengue pooled convalescent serum, likewise shows a similar pattern, as seen in Panel B. In Panel C, our ADE model was used to simulate {\it in vitro} experiments in which antibody level is kept constant and infection load is measured after 2 days. Qualitatively all three panels show the same pattern: lower levels of antibody increased viral growth but higher levels were neutralizing. (West Nile data digitized from Figure 3A in ~\cite{pierson2007stoichiometry}. Dengue data digitized from Figure 4A  in~\cite{dejnirattisai2010cross}.)}
\label{fig2}
\end{figure}

In this section, we simulate a 48 hour {\it in vitro} cell culture experiment. As there is no active immune response, $A_c(0)=A_n(0)=0$. Fig~\ref{fig2} shows the quantity of infected cells at the end of the simulation for different values of $A_p$ and compares them to experimental values.

An antibody level of 15 produces maximum ADE in the model; this level of antibody produces 10 fold greater infection load after 48 hours as compared to no antibody. Hence the model described above does produce ADE. Higher levels of antibody ($>74$) reduce infection growth as compared to no antibody. This pattern of enhanced growth at intermediate levels of antibody and reduced or no growth at higher levels of antibody matches the pattern seen from {\it in vitro} experiments with dengue ~\cite{beltramello2010human} and West Nile ~\cite{pierson2007stoichiometry} viruses.
\subsection*{The model produces EI-PA}

\begin{figure}[!h]
\centering
  \includegraphics[scale=0.75]{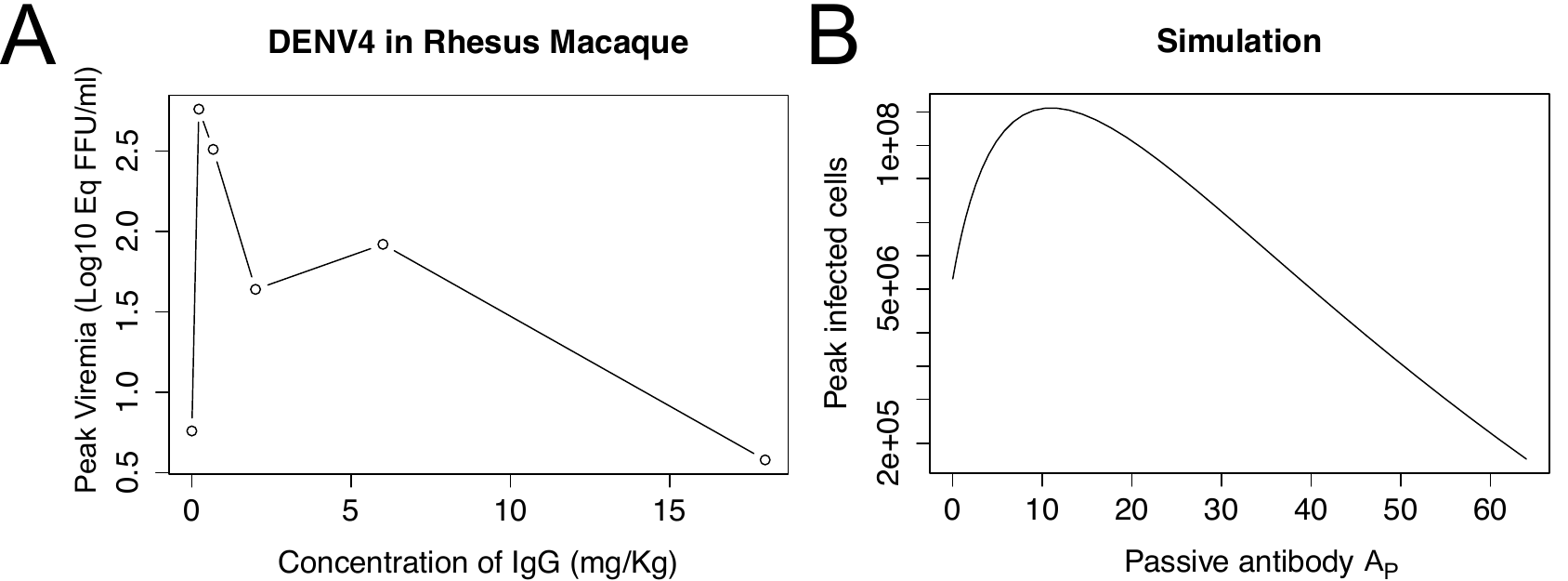}
\caption{{\bf  Passive antibody can enhance infection.}
The figure shows experimental data, Panel A, and simulation results from the ADE model in Panel B. Adding lower levels of passive antibody (e.g. maternal antibody) enhances infection, but higher levels of passive antibody are protective. (Experimental data from Table 1 in ~\cite{goncalvez2007monoclonal}.)}
\label{fig3}
\end{figure}

Building upon our previous {\it in vitro} model to replicate an {\it in vivo} experiment, we now include additional factors.
First, we simulate supply of passive (e.g. maternal) antibody ($A_p$) to an infected host. Second, in addition to these passive antibodies, the host’s immune system will generate antibodies ($A_n$) in response to the infection. Fig~\ref{fig3} shows peak infected cells for different values of $A_p$.

In this case, a passive antibody level of 11 produces maximum enhancement; this level of passive antibody increases infection load 35 fold compared to no passive antibody. Thus, in this model, adding passive antibody can dramatically enhance infection, but higher levels of passive antibody are protective. Hence, the model qualitatively reproduces dengue data from infants ~\cite{kliks1988evidence} ~\cite{chau2008dengue} which show that high levels of maternal antibody are protective but intermediate levels are associated with enhanced disease. The model is also consistent with dengue animal studies showing enhanced viral load when passive antibody is supplied ~\cite{halstead1979vivo} ~\cite{goncalvez2007monoclonal}.

\subsection*{The model does not (necessarily) produce EI-HM}

\begin{figure}[!h]
\centering
  \includegraphics[scale=0.5]{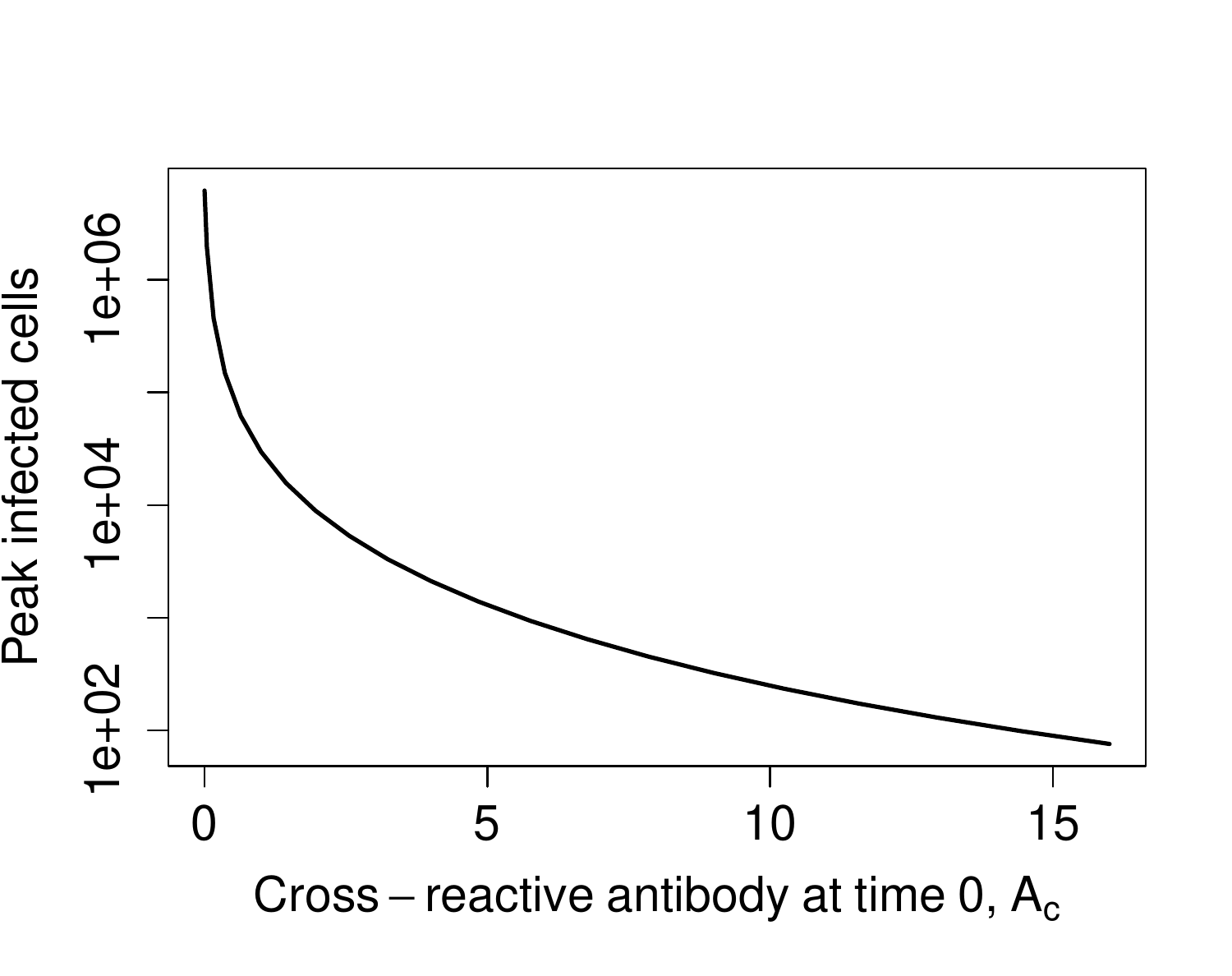}
\caption{{\bf ADE but no EI-HM.}
 The figure shows simulation results from the ADE model. In these simulations, cross-reactive antibodies behave the same as {\em de novo} antibodies ($s_c=s_n$ and $k_{ci} = k_{ni}$). Boosting the baseline level of cross-reactive humoral immunity does not enhance infection in these simulations despite the presence of ADE.}
\label{fig4}
\end{figure}

In this section, we simulate infection for varying baseline levels of cross-reactive memory antibody. We build upon our original {\it in vitro} model and now include both {\it de novo} and cross-reactive antibody response to create an {\it in vivo} model. Because cross-reactive antibodies may replicate some but not necessarily all the features of {\it de novo} antibodies such as growth rate or neutralizing activities, we first investigate the simplest scenario where cross-reactive and {\it de novo} antibodies have all of the same properties. Fig~\ref{fig4} shows peak infected cells for different values of $A_c(0)$.

In this case, we see only protection as $A_c(0)$ increases. Hence ADE, where partially bound virus is more infectious than free virus, does not necessarily produce EI-HM.

\subsection*{The model can produce EI-HM in certain circumstances}

Although the simplest  implementation of our {\it in vivo} model (shown above) does not show enhancement of infection from humoral memory, our model produces EI-HM in certain situations as shown in the following sections.

\subsubsection*{Cross-reactive antibody is less neutralizing}
\begin{figure}[!h]
\centering
  \includegraphics[scale=0.5]{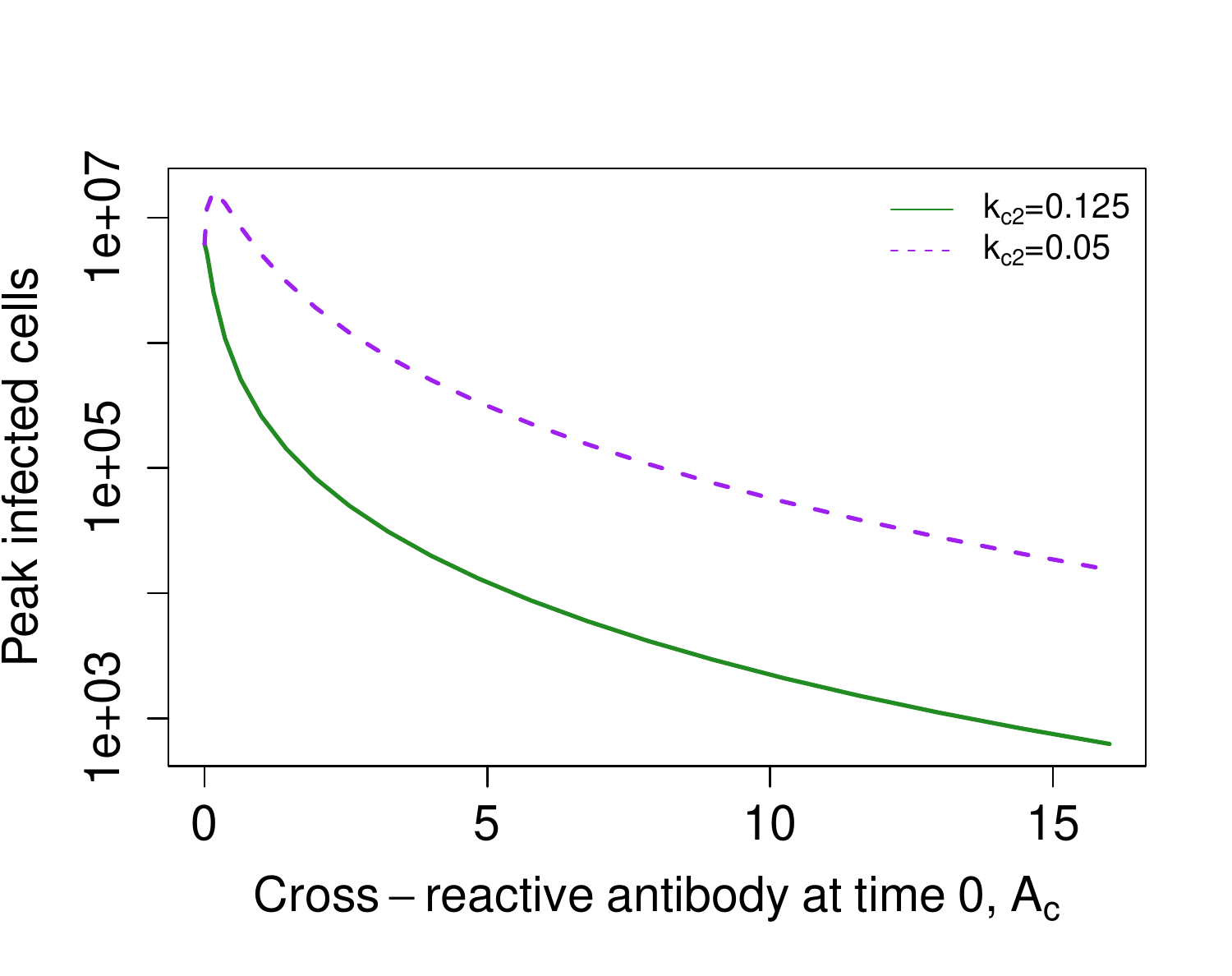}
\caption{{\bf Less neutralizing, cross-reactive antibodies, modeled with $\mathbf{k_{c2}=0.05}$, produce some EI-HM.}
 The figure shows simulation results from the ADE model when including cross-reactive antibodies. When the cross-reactive antibody is half as neutralizing as the {\it de novo} antibody ($k_{c2}=0.125$), there is no EI-HM. But, when the cross-reactive antibody is 5 times less neutralizing than the {\it de novo} antibody ($k_{c2}=0.05$), there is some enhancement of infection at low levels of baseline cross-reactive antibody.}
\label{fig6}
\end{figure}

In these simulations the cross-reactive antibody grows at the same rate but is less neutralizing than the {\it de novo} antibody – either $k_{c2}=0.125$ or $k_{c2}=0.05$. $k_{c2}=0.125$ can be loosely interpreted as meaning that the cross-reactive antibody is half as neutralizing as the {\it de novo}  antibody, whereas $k_{c2}=0.05$ can be loosely interpreted as meaning that the cross-reactive antibody is 5 times less neutralizing than the {\it de novo}  antibody. Fig~\ref{fig6} shows peak infected cells for different values of $A_c(0)$.

With $k_{c2}=0.125$, no EI-HM was observed, but with $k_{c2}=0.05$, there is some enhancement of infection at low levels of $A_c(0)$. However, the degree of enhancement is again relatively small – a 5 fold difference in neutralization ability translates into a maximum enhancement of only 2.7 fold compared to $A_c(0)=0$.

Increasing $k_{c1}$, the rate at which the cross-reactive antibody enhances virus, has an analogous effect to decreasing $k_{c2}$. $k_{c1}=2k_{n1}=2$ does not produce EI-HM whereas $k_{c1}=5k_{n1}=5$ produces maximum relative enhancement of 2.7 fold.
\subsubsection*{Non-antibody immune response is dominant}
\begin{figure}[!h]
\centering
  \includegraphics[scale=0.8]{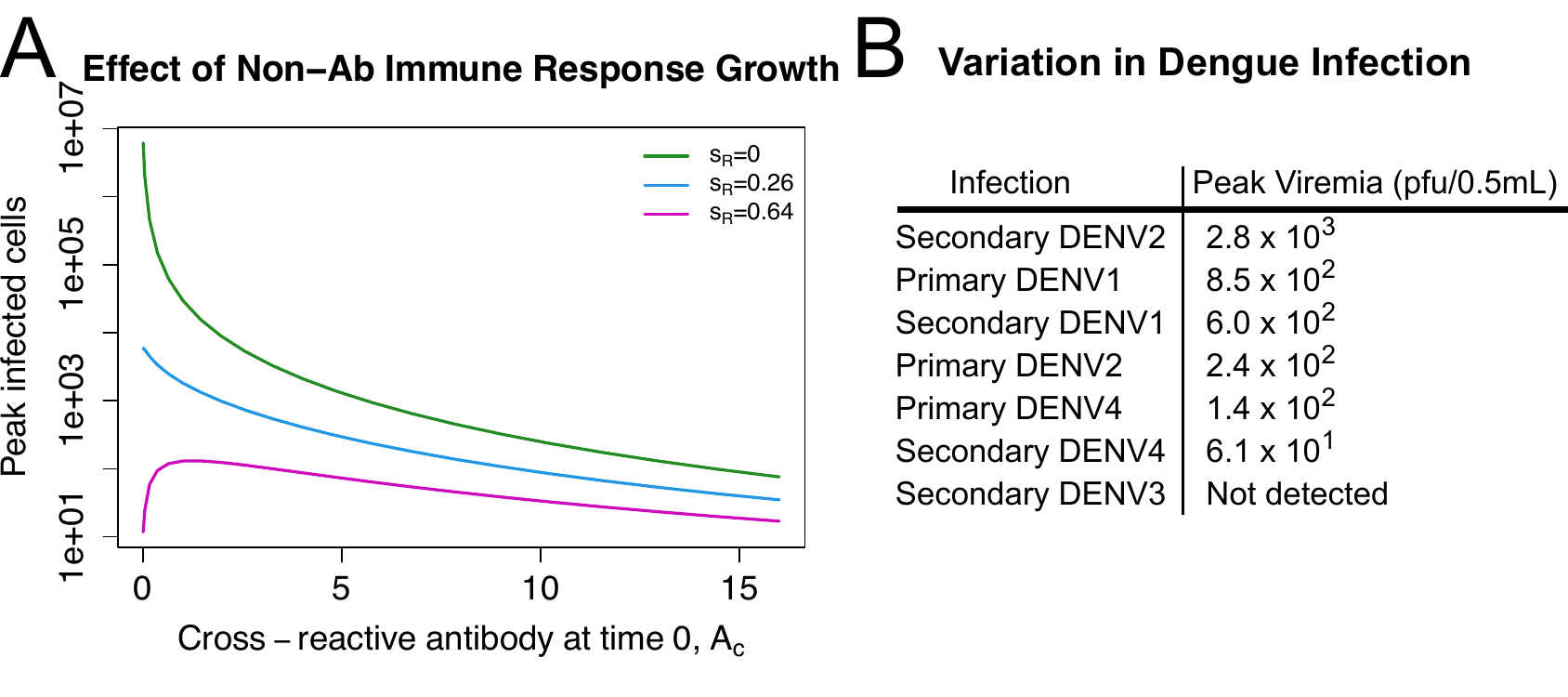}
\caption{{\bf High growth rates of the non-antibody immune response, such as $\mathbf{s_R=0.64}$,  can produce dramatic EI-HM.}
Panel A shows simulation results from the ADE model. When the non-antibody immune response is  strong ($s_R=0.64$), controlling infection several days before antibody levels become neutralizing, there is the possibility of dramatic EI-HM of up to 11 fold.  In contrast when $s_R \leq0.26$,  there is no EI-HM. Peak viremia has a greatly varying range from experimental challenges with dengue in rhesus macaques, as shown in Panel B. In these experiments peak viremia was greatly increased in secondary DENV2 infections but below detection in secondary DENV3. Both the simulation and the experimental data suggest the possibility of dramatic EI-HM but also its inconsistency. Experimental data from ~\cite{halstead1973studies}}
\label{fig7}
\end{figure}

In these simulations we vary the growth rate, $s_R$, of the non-antibody immune response. To compare, $s_R=0$ shows the peak of infected cells when there is no non-antibody immune response and is therefore equivalent to the results in Fig~\ref{fig4}. Fig~\ref{fig7} shows peak infected cells for different values of $A_c(0)$.

A growth rate of $s_R=0.26$ is the highest value of $s_R$ that shows no EI-HM whereas $s_R=0.64$ corresponds to primary infection peaking at day 5 (see Fig S7), which is many days before antibody reaches sterilizing levels. With $s_R=0.64$, enhancement of up to 11 fold is possible. However, the range of $A_c(0)$ values that lead to enhancement is quite a bit smaller than observed {\it in vitro},  $>0$ to 74 compared to $>0$ to 24.

\subsection*{Other models}
We consider three changes to our overarching ADE model: 1) B cells and plasma cells are explicitly included in the model, 2) step functions in the differential equations are replaced with Hill functions, and 3) dissociation between antibody and virus is explicitly modeled. See Supplement for a more detailed description of these changes. None of these changes qualitatively change our results (figures not shown).

Explicitly modeling the dissociation between antibody and virus also allows us to consider the situation where the cross-reactive antibody has a higher dissociation rate than the {\it de novo} antibody. This situation is similar to that described in § Cross-reactive antibody is less neutralizing. When the dissociation rate of cross-reactive antibody is 50/day versus 10/day for the {\it de novo} antibody, no EI-HM is observed. But when the dissociation rate of cross-reactive antibody is 150/day versus 10/day for the {\it de novo} antibody, there is some EI-HM at low levels of $A_c(0)$ but with maximum enhancement of 1.1 fold. See \nameref{S1 Text} for more details.

\subsection*{Suppressive memory can produce EI-HM without ADE}

All of our above simulations show memory dominance when $A_c(0)$ is sufficiently large – $A_c(0)>0.036$. This shows that memory dominance does not necessarily mean EI-HM. In fact, with our model ADE ($\beta_2 >\beta_1$) is necessary – but not sufficient – to produce EI-HM.

\begin{figure}[!h]
\centering
  \includegraphics[scale=0.5]{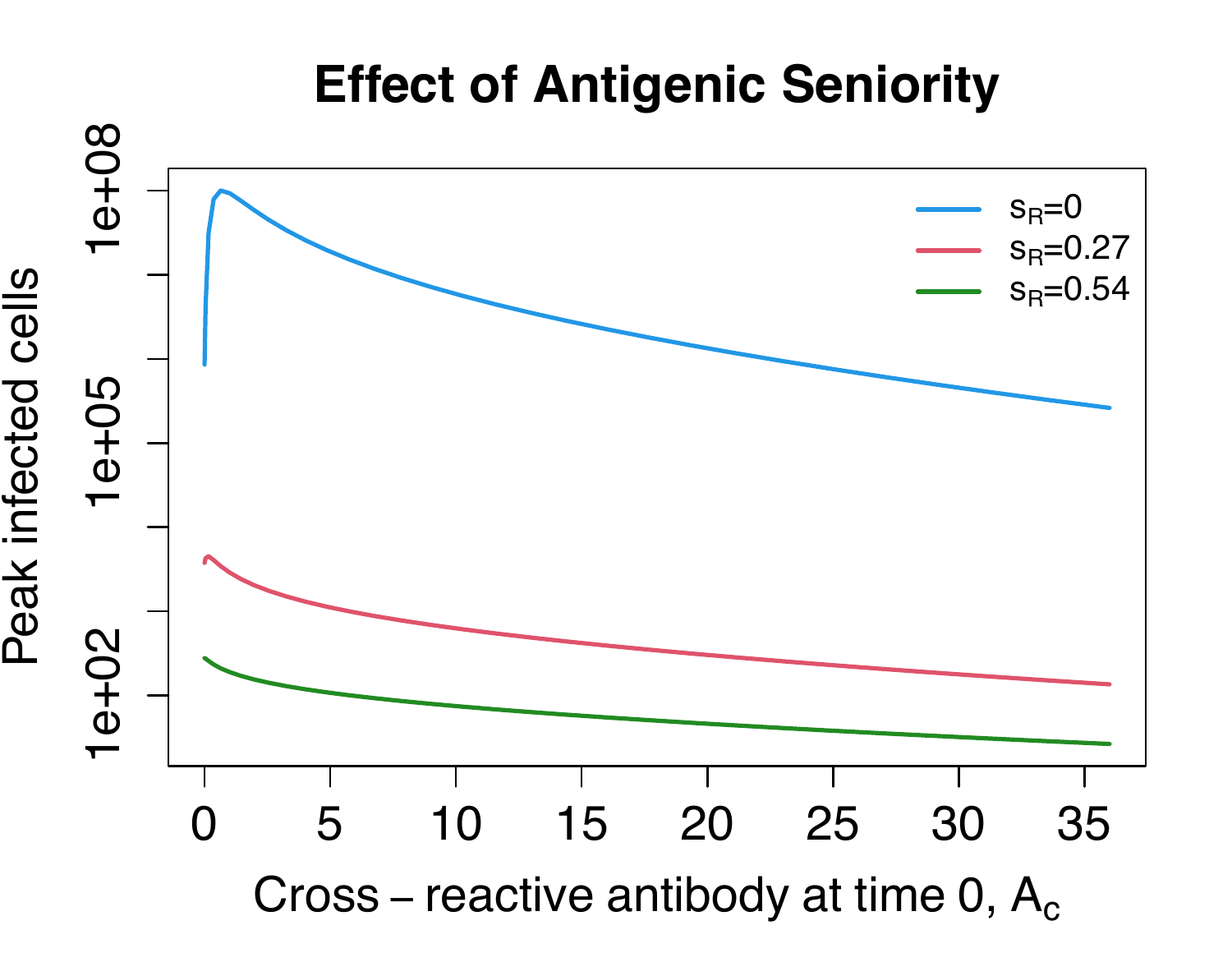}
\caption{{\bf Suppressive memory can produce EI-HM even in the absence of ADE. }
The figure shows simulation results from the model described in § Suppressive memory can produce EI-HM without ADE. In this case the combination of suppressive memory and lower potency of the cross-reactive antibody leads to EI-HM at lower levels of baseline cross-reactive antibody even though there is no ADE. The presence of non-antibody immune responses ($s_R >0$) can lessen or even prevent this effect.}
\label{fig8}
\end{figure}

Our ADE model does not incorporate suppressive memory. In this section we modify the equations for $A_c$ and $A_n$ to incorporate this effect.
\begin{equation}
\dot{A_c}= s_c\chi(I>1)\frac{\phi A_c}{\phi+A_c+A_n}
\end{equation}
\begin{equation}
\dot{A_n}= s_n\chi(I>1)\frac{\phi A_n}{\phi+A_c+A_n}
\end{equation}
Here $\chi$ is the indicator function, $\phi=28$, and $s_c=s_n=1.5$. In this system increasing $A_c$ suppresses the growth rate of $A_n$ and vice versa. We consider the situation where $A_c$ is a quarter as potent as $A_n$ ($k_{c1} =0.25, k_{c2}=0.0625$) and there is no ADE ($\beta_1=\beta_2=40$). In this case there is dramatic EI-HM with fold enhancement of up to 117 fold (Fig~\ref{fig8}). However, adding a non-antibody immune response with $s_R=0.27$ or $s_R=0.54$ mostly or completely abolishes EI-HM. Here $s_R=0.54$ corresponds to primary infection peaking at day 5 (Fig S8).

\subsection*{Other adverse effects of humoral memory}
EI-HM is one possible adverse effect of humoral memory. Another possible adverse effect is a negative effect of humoral memory at baseline on humoral memory post infection. Such an effect is possible because of the negative feedback between antibody and pathogen and does not require ADE or suppressive memory (see ~\nameref{S10_Fig}). Hence, if a class of antigenically similar pathogens can infect more than once, humoral memory may protect against the immediate infection but enhance later infection.

\section*{Discussion and Conclusion}
\subsection*{Intuitive Explanation}
While our results may seem disparate, they can be explained by two factors: 1) the effect of antibody early in infection and 2) the effect of cross-reactive humoral memory on the rapidity of humoral response.

In the case of ADE, antibody may enhance viral growth early in infection, but humoral memory may also accelerate the development of protective levels of antibody. In the case of suppressive memory, viral growth is reduced early in infection, but the development of a potent antibody response may also be delayed. 

Depending on which of these factors dominates, cross-reactive humoral memory may either enhance or decrease infection. 
\subsection*{Concluding Remarks}
Our results are consistent with the hypothesis that ADE and enhancement of infection from passive antibody (EI-PA) can result when virus with intermediate levels of antibody is more infectious than virus with lower or higher levels of antibody ~\cite{pierson2007stoichiometry}. Hence additional mechanisms, such as suppression of the immune response ~\cite{suhrbier2003suppression}, are possible but not necessary to produce the phenomena of ADE and EI-PA.

More importantly we show that ADE, EI-PA, and enhancement of infection from cross-reactive humoral memory (EI-HM) are distinct. ADE in our models always implies the possibility of EI-PA. Furthermore, the range of antibody levels that produces ADE is similar to the range of passive antibody levels that produces EI-PA. This is because passive antibody can enhance viral growth early in infection but does very little to accelerate the development of protective levels of antibody. In contrast EI-HM is a double-edged sword, and ADE and EI-PA may not imply EI-HM. 

In our models, when ADE does produce EI-HM the range of baseline antibody levels that produce EI-HM is very different from the range that produces ADE. This difference in ranges may explain the observation in dengue patients that ADE measurements of pre-infection plasma did not positively correlate with peak viremia ~\cite{laoprasopwattana2005dengue}.

ADE, EI-PA, and EI-HM are of particular concern for dengue infection. In the case of dengue, there is extensive evidence for ADE {\it in vitro}  ~\cite{beltramello2010human} ~\cite{dejnirattisai2010cross} ~\cite{rodenhuis2010immature} ~\cite{schieffelin2010neutralizing} ~\cite{flipse2016antibody}, and EI-PA of up to 100 fold has been demonstrated in monkey experiments ~\cite{halstead1979vivo} ~\cite{goncalvez2007monoclonal}. In contrast, the evidence for greater infection load in secondary infection is more limited. A large monkey experiment involving 118 rhesus macaques showed higher viral load in secondary DENV2 infection as compared to primary DENV2 infection. However, in the same study secondary infections with DENV1, DENV3 and DENV4 showed reduced viremia when compared to the respective primary infections ~\cite{halstead1973studies}. A recent epidemiological study showed increased risk of severe dengue in children with low or intermediate levels of dengue antibody at baseline as compared to dengue seronegative children; higher levels of baseline antibody were associated with reduced risk. In contrast, risk of symptomatic dengue decreased essentially monotonically with antibody levels ~\cite{katzelnick2017antibody}. Likewise analysis of clinical trials data suggested that vaccination of dengue naive children with the CYD-TDV dengue vaccine increased the risk of severe dengue but reduced the risk of symptomatic dengue \cite{sridhar}. Our results suggest that this pattern could be explained by a scenario in which the majority do not experience EI-HM but a minority – for example those with very poor quality cross-reactive antibody – do experience EI-HM and this minority also tends to have more severe disease. However, ~\cite{katzelnick2017antibody} and\cite{sridhar} examined disease severity rather than infection load, and in ~\cite{katzelnick2017antibody} the degree of enhancement was highly sensitive to the method of classifying disease severity. (Using the 1997 World Health Organization (WHO) dengue severity classification system, the risk of severe dengue was up to 7.6 fold elevated. Using the newer 2009 WHO severity classification, generally regarded as more accurate ~\cite{horstick2015dengue} ~\cite{guzman2016dengue}, this was reduced to only 1.75 fold.) Despite this limitation, in the case of human dengue infections, our results support the hypothesis that ADE sometimes produces EI-HM but often does not.

We show that, in principle, memory dominance plus suppressive memory can also produce EI-HM. In contrast memory dominance alone does not produce EI-HM in any of our models. However, inclusion of a non-antibody immune response in our simulations dramatically reduces EI-HM from suppressive memory, which casts doubt on its real world relevance.  

It should be emphasized that our simulations only consider the effect of humoral memory present at baseline on that particular infection. A negative effect of humoral memory at baseline on humoral memory post infection can occur as a consequence of the negative feedback between antibody and pathogen and does not require ADE or suppressive memory.  For infections, like influenza, where reinfections are common, some of the apparent adverse effects of humoral memory likely involve not EI-HM but the negative effect of humoral memory at an earlier time point on humoral memory at a latter time point with potentially many infections in between ~\cite{linderman}.

Although CD4 T cells can increase pathology ~\cite{macmaster}, a mechanism such as ADE is not established for T cells. This implies that activating non-antibody immune responses, especially CD8 T cells, could be an important overlooked component in vaccine development. Although memory dominance and suppressive memory are concerns for antibody as well as T cells, our results suggest that suppressive memory is unlikely to cause EI-HM. Moreover, the ability to strategically direct T cell responses has been demonstrated ~\cite{vezys}~\cite{walsh}, potentially turning memory dominance into a long term advantage.

Finally, our results also illustrate potential pitfalls from mathematical modeling, which may be relevant not only for biology but also for other disciplines, such as economics and epidemiology, that use abstract mathematical models. In the simplest version of our models, there is EI-HM from suppressive memory but none from ADE, whereas adding plausible complexity reverses this situation.

\section*{Supplement}

\paragraph*{S1 Text.}
\label{S1 Text}
{\bf Model Dynamics and Alternative Model Results.} 
\paragraph*{Fig S1.}
\label{S1_Fig}
{\bf  Model Schematic.}
In our model, infected cells ($I$) produce free virus($V_1$) and can be removed via cell death or the non-antibody immune response ($R$). Virus can be successively bound with only free ($V_1$) and partially bound virus ($V_2$) able to infect cells. Binding can come from the {\it de novo} antibody response ($A_n$), the cross-reactive antibody response ($A_c$), or the passive antibody response ($A_p$), with $A_n$ and $A_c$ growing in response to the infected cells.
\paragraph*{S2 Fig.}
\label{S2_Fig}
{\bf  Dynamics of infection under ADE  \textbf{\textit{in vitro}}.}
In this {\it in vitro} situation, where there is no growth in immune response, the phenomena of ADE is evident as the number of infected cells after 48 hours is greater at intermediate ranges of antibody. We show the differences in dynamics for three different initial antibody concentrations in Panels B and C: low but non-zero (black), intermediate (red), and high (blue). 
\paragraph*{S3 Fig.}
\label{S3_Fig}
{\bf  Dynamics of infection given passive antibodies.}
When passive antibodies are given and a {\it de novo} antibody response is mounted, again ADE appears at intermediate levels (red). 

\paragraph*{S4 Fig.}
\label{S4_Fig}
{\bf  Dynamics of infection given cross-reactive antibodies does not necessarily produce enhancement of infection in these simulations where \textbf{\textit{de novo}} and cross-reactive antibodies are identical.}
When there are both {\it de novo} and cross-reactive antibodies present and growing and binding at the same rate, enhancement of infection is not seen, as shown in Panel A; rather, the greater initial cross-reactive antibody load the lower the viral load and number of infected cells as seen in Panels B-C. Here, black represents low initial cross-reactive antibody concentration, red intermediate, and blue high.
\paragraph*{S5 Fig.}
\label{S6_Fig}
{\bf  Less neutralizing cross-reactive antibodies can sometimes cause EI-HM.} If cross-reactive antibodies are approximately a fifth as good as {\it de novo} antibodies, some EI-HM may occur as shown in Panel A with slightly higher peak viremia and infected cell counts. While the the line for the low cross-reactive antibody concentration is the same for both $k_{c2}$ values, as seen in black in Panels B and C, they differ for the intermediate (red) and high (blue) concentrations where $k_{c2} = 0.125$ is given by the solid lines and $k_{c2} = 0.05$ by the dashed lines.
\paragraph*{S6 Fig.}
\label{S6_Fig}
{\bf  Dynamics of infection given cross-reactive antibodies and non-antibody immune response.} Depending on the rate of growth of non-antibody immune response and initial cross-reactive antibody amount, EI-HM can sometimes occur and give a 10-fold increase in peak infected cells. Colors denote level of initial cross-reactive antibody with black, red, and blue representing low, intermediate, and high values. Line types denote non-Ab immune response growth rates with solid, dashed, and x'd lines representing $s_R=0$, $s_R= 0.26$, and $s_R=0.64$ respectively.
\paragraph*{S7 Fig.}
\label{S7_Fig}
{\bf  Dynamics of infection given suppressive memory.}  
As shown in Panel A given the scenario where there is no non-antibody immune response, EI-HM can occur with a fold change of $117$ for $s_R=0$. With non-antibody immune responses, however, enhancement of infection is greatly reduced with a  maximum fold change of $1.2$ for $s_R=0.27$ or nonexistent for $s_R = 0.54$. In Panel B, we consider the viral dynamics for low (black), intermediate (red), and high (blue) points for each of the different growth rates where $s_R=0$ is a solid line, $s_R=0.27$ is dashed, and $s_R=0.54$ is x'd. Note that for $s_R=0$ in particular if the high point was taken at an even higher antibody level ($>25$), it would in fact peak below the low point. 
\paragraph*{S8 Fig.}
\label{S9_Fig}
{\bf  $\mathbf{k_{\text{off,c}}=100/}$day produces some EI-HM.}
 The figure shows simulation results from a modification to the simple ADE model such that dissociation between antibody and virus is explicitly modeled. When the cross reactive antibody has 5 times the dissociation rate of the de novo antibody ($k_{\text{off,c}}=50/\text{day}$ versus $k_{\text{off,n}}=10/\text{day}$), there is no EI-HM. But, when the cross reactive antibody has 10 times  or more the dissociation rate of the de novo antibody ($k_{\text{off,c}}=100/\text{day}$ or $k_{\text{off,c}}=150/\text{day}$ and $k_{\text{off,n}}=10/\text{day}$), there is some enhancement of infection at low levels of baseline cross reactive antibody.

\paragraph*{S9 Fig.}
\label{S10_Fig}
{\bf  Effect of baseline humoral memory on postinfection antibody levels..}
In this model, without ADE or suppressive memory, and with non-antibody immune response we see that higher levels of baseline antibody ($>$5.8) negatively affect final (post infection) antibody levels.


\nolinenumbers

%
%
%

\end{document}